\documentclass[preprint]{revtex4}
\usepackage{amsfonts}
\usepackage{amsmath}
\usepackage{amssymb}
\usepackage{graphicx}

\begin{document}

\title{\bf Search for anomalous single production of the fourth SM 
family quark decaying into a light scalar} 

\author{E. Arik$^{a}$, O. \c Cak\i r$^{b}$, S. Sultansoy$^{c,d}$} 
\affiliation{$a$) Bo\u{g}azi\c{c}i University, Faculty of Arts and Sciences, 
Department of Physics,   80815, Bebek, Istanbul, Turkey \\
$b$) Ankara University, Faculty of Science, 
Department of Physics, 06100 Tando\u gan, Ankara, Turkey \\
$c$) Gazi University, Faculty of Arts and  Sciences, 
Department of Physics, 06500, Teknikokullar, Ankara, Turkey \\
$d$) Azerbaijan Academy of Sciences, 
Institute of Physics, H. Cavid Av., 33, Baku, Azerbaijan }

\begin{abstract}
Superjet events observed by the CDF Collaboration are interpreted as 
anomalous single production of the fourth SM family  $u_4$ quark, decaying 
into a new light scalar particle. The specific predictions of the proposed 
mechanism are discussed.

\end{abstract}
\maketitle
\vskip 1.0cm

Recently, CDF Collaboration has reported {\cite{1}} the observation of an 
excess of events in $W + 2 , 3$ jet topologies, the so called ``superjet'' 
events, in which a single jet has both a soft lepton and a secondary vertex. 
These events are interpreted in {\cite{2}} as scalar quark $\tilde b$ 
with a mass of $3.6$ GeV and a lifetime of $1$ ps, decaying into 
$c \, l \tilde \nu$ where scalar neutrino is assumed to be massless. 
In our opinion, this interpretation favors the MSSM scenario 
with superpartner of the right-handed neutrino as the LSP {\cite{3}}, 
because LEP experiments put the limit $m_{\tilde \nu} > 44$ GeV on the 
mass of superpartners of the left-handed neutrinos {\cite{4}}. 
As mentioned in {\cite{5}}, at least some of these events could be explained 
as anomalous single production of the fourth SM family (see {\cite{saleh}} 
and the references therein) quarks via the 
process  $u_4 \to tg \to W b g$. The superjet is associated with the normal 
$b$ quark decay. However, 
this interpretation leads to an essential excess in single tag events, 
which is several times larger than CDF observations {\cite{1}}. 

In this letter, we give another interpretation to superjet events. 
Following our previous study {\cite{5}}, we still assume that the production 
is due to anomalous $q_4qg$ interaction but the decay chain is changed to  
$u_4 \to t x^0 \to W  b \, \tau^+ \tau^-$. The new proposed particle 
$x^0$ must have a mass around  $4-5$ GeV. In this mechanism, superjet 
is formed by the decay of one $\tau$ leptonically and the other one 
hadronically. There are two  possible identifications of $x^0$. One 
possibility is the lightest neutral Higgs boson in SM with extended 
Higgs sector. The other one is a light neutral dilepton which is expected 
in preonic models. 

The effective Lagrangian for the  anomalous interactions between the  
fourth family quarks, ordinary quarks and the gauge bosons $V$ 
($V = \gamma, Z, g$) can be written as follows:
\begin{eqnarray}
L = \frac {\kappa_{\gamma}^{q_i}}{\Lambda} e_q g_e \bar q_4 \sigma_{\mu \nu} (A_{\gamma}^{q_i} + B_{\gamma}^{q_i} \gamma_5) q_i F^{\mu \nu} + 
 \frac {\kappa_Z^{q_i}}{2 \Lambda} g_Z \bar q_4 \sigma_{\mu \nu} 
(A_Z^{q_i}  + B_Z^{q_i}  \gamma_5) q_i Z^{\mu \nu} \nonumber \\
+\,  \frac {\kappa_g^{q_i}}{\Lambda} g_s \bar q_4 \sigma_{\mu \nu} 
(A_g^{q_i}  + B_g^{q_i}  \gamma_5) T^a q_i G^{\mu \nu}_a + h.c. 
\end{eqnarray}  
where $F^{\mu \nu}, Z^{\mu \nu}$, and $G^{\mu \nu}$ are the field strength 
tensors of the photon, $Z$ boson and gluons, respectively; $T^a$ are 
Gell-Mann matrices;
$e_q$ is the charge of the quark; 
$g_e, g_Z$, and $g_s$ are the electroweak, and 
the strong coupling constants respectively.  
$g_Z = g_e/\cos\theta_W \sin\theta_W$ where $\theta_W$ is the Weinberg angle. 
 $A_{\gamma, Z, g}^q$ and 
$B_{\gamma, Z, g}^q$ are the magnitudes of the neutral currents; 
$\kappa_{\gamma, Z, g}$ define the strength of the  anomalous couplings for the neutral currents with a photon, a $Z$ boson 
and a gluon, respectively;  $\Lambda$ is the cutoff scale for the new 
physics.
We assume all the neutral current magnitudes in Eq. (1) to be equal, 
satisfying the constraint $|A|^2 + |B|^2 = 1$ and take all anomalous 
couplings as $\kappa_{\gamma}^{q_i} =  {\kappa_Z^{q_i}} = {\kappa_g^{q_i}}  
= 1 $. We have implemented the new 
interaction vertices into the CALCHEP {\cite{calchep}} package to calculate 
the cross sections and branching ratios. We use the parton distribution function CTEQ5M \cite{7_1} 
and the scale $Q=m_{u_{4}}$.  

{\it{$x^0$ as a light Higgs}}. We require
 $u_4tx^0$ coupling to be sufficiently large in order to provide significant 
BR($u_4 \to t x^0$). On the other hand, the interaction of $x^0$ with the 
intermediate vector bosons should be suppressed in order to avoid 
contradiction with the LEP data. So, we will denote this particle as 
a light Higgs boson $h^0$. 
Naturally, the main decay mode of $h^0$ should be into kinematically 
allowed heaviest fermions, namely $\tau^+ \tau^-$ and $c \, \bar c$. 
In the case of  $h^0 \to \tau^+ \tau^-$, the branching ratios into possible 
final states are 0.44 for superjet,  0.12 for all lepton mode, 0.43 for 
all hadron mode. Similar numbers are valid for $h^0 \to c \, \bar c$.   
In order to have the lifetime of $h^0$ to be less than $1$ ps,  the 
coupling constant $a_{\tau}$  in the interaction term 
$a_{\tau}h^0 {\bar\tau \tau}$ 
 should be larger than $5 \times 10^{-7}$. 

The number of superjet events can be estimated as 
\begin{eqnarray}  
N_{s} &=& \sigma(gq \to u_4)\cdot BR(u_4 \to t h^0) \cdot 
BR(t \to b W)   \nonumber \\
&& \cdot BR(h^0 \to superjet) \cdot  BR(W \to e \nu + \mu \nu) \cdot 
\epsilon \cdot L_{int}
\end{eqnarray}
where $\epsilon$ stands for detection efficiency. 
In Table 1, we present production cross section,  branching ratios and total 
decay width for $u_4$ at different mass values, assuming 
BR($u_4 \to t \, h^0$) $= 10\%$. This assumption corresponds to $b_{t u_4} 
\approx 0.1$ where $b_{t u_4}$ is the coupling in the interaction term 
$b_{t u_4}h^0 {\bar t} u_4$. By taking $\epsilon = 0.25$, 
BR($t \to b W$) $= 1$ and BR($W \to  e \nu + \mu \nu $) $= 0.21$, we obtain 
from Eq. (2) number of superjet events $N_{s} = 6$ for $m_{u_4} = 300$ GeV 
and the integrated luminosity $L_{int} = 106 
$ pb$^{-1}$.

\begin{table}[tbp]
\begin{center}
\caption{Branching ratios ($\%$), total decay widths for $u_4$ and production 
cross section $\sigma(p \bar p \to u_4 X)$ with $\Lambda = 2$ 
TeV.}
\vskip 0.1 cm
\begin{tabular}{|c|c|c|c|c|c|c|c|c|}
\hline
Mass (GeV)& $gu(c)$ & $gt$ & $Zu(c)$ & $Zt$ & $\gamma u(c)$ &  $\gamma t$ & 
$\Gamma$ (GeV)  & $\sigma$ (pb) \\
\hline
200 & 42 & 0.54 & 2.0 & - & 0.90 & 0.028 & 0.39 & 33.4 \\
\hline
250 & 40 & 5.2 & 2.2 & - & 0.83 & 0.11 & 0.81 & 25.9 \\
\hline
300 & 37 & 11 & 2.2 & 0.41 & 0.77 & 0.23 & 1.50 & 24.8 \\
\hline
400 & 32 & 17 & 2.1 & 1.0 & 0.69 & 0.37 & 3.96 & 21.8 \\
\hline
500 & 31 & 21 & 2.0 & 1.4 & 0.66 & 0.44 & 8.20 & 15.2 \\
\hline
600 & 30 & 23 & 2.0 & 1.5 & 0.64 & 0.49 & 14.67 & 9.4 \\
\hline
700 & 30 & 24 & 2.0 & 1.6 & 0.62 & 0.51 & 23.80 & 5.1 \\
\hline
\end{tabular}
\end{center}
\end{table}

{\it{$x^0$ as a light dilepton}}. Second candidate for the identification of 
$x^0$ is a neutral dilepton with zero lepton number {\cite{7}}. So, we 
denote this particle as $D^0_{\tau}$. The difference between  $D^0_{\tau}$ 
and  $h^0$ decay modes is that  $D^0_{\tau}$ decays into  $\tau^+ \tau^-$ 
only.  $D^0_{\tau}$ can be produced in $u_4$ decays via the mixing with a 
diquark   $D^0_q$, interacting with $u_4$ and $t$ (for classification of 
diquarks see {\cite{8}}): 
\begin{eqnarray}
D^0_1 = D^0_{\tau} \cdot \cos\theta +  D^0_q \cdot \sin\theta \\
D^0_2 = - D^0_{\tau} \cdot \sin\theta +  D^0_q \cdot \cos\theta
\end{eqnarray}
where $\theta$ is the mixing angle.
Therefore, $u_4$ decay chain becomes $u_4 \to t D^0_1 
\to W  b \, \tau^+ \tau^-$, where we identify  $x^0$ as  $D^0_1$, and  
 $D^0_2$ is assumed to be heavy. In order to obtain BR($u_4 \to t \, D^0_1$)$ 
= 10\%$, one needs $b \sin\theta \, = 0.1$ where $b$ is the coupling 
constant of the $D^0_q u_4 t$ interaction.

To test the proposed mechanism, we have 
reconstructed the invariant mass of the  $W + j$ system where $j$ denotes 
the ordinary jet accompanying the superjet, using the W + 2 jet events in 
Table XVI of {\cite{1}}. The resulting mass values are 
$159, 175, 600, 159, 178, 228, 172,$  $149$ GeV. The third and the sixth 
events  do not reproduce the expected top quark mass value and probably 
they are due to SM background. 

Another prediction of our mechanism is the occurrence of  spectacular events 
with collinear $e \mu $ tracks originating from $x^0$  when both $\tau$ 
leptons 
decay leptonically. Ratio of the number of events of this type to the number 
of superjet events should be $1:15$ for $h^0$ and $1:8$ for $D_{\tau}^0$. 
Therefore, we expect the observation of a number of  collinear $e \mu$ tracks 
at the upgraded Tevatron with integral luminosity $1$ fb$^{-1}$.

In conclusion, CDF superjet events seem to indicate very exciting physics 
at TeV scale. Namely, existence of the fourth SM family, anomalous 
interactions, extended SM Higgs sector etc. The scale $\Lambda = 2$ TeV 
can be a hint of relatively low scale compositeness which will lead to a 
rich {\it {zoo}} of new particles at the LHC.

\begin{center}{\bf Acknowledgments} \\
\end{center} 
We are grateful to P. Giromini for his illuminating remarks on CDF data.  
This work is partially supported by Turkish State Planning Organization 
under the Grant No 2002K120250.

\end{document}